\RequirePackage{fix-cm}
\documentclass[smallcondensed]{svjour3}     % onecolumn (ditto)
\smartqed  % flush right qed marks, e.g. at end of proof
\usepackage{graphicx}
\usepackage{color}
\usepackage{booktabs}
\usepackage{upgreek}
\usepackage[utf8]{inputenc}
\usepackage{lpic}
\journalname{Hyperfine Interactions}
\begin{document}

\title{Towards Measuring the Ground State Hyperfine Splitting of Antihydrogen -- A Progress Report\thanks{European Research Council under European Union’s Seventh Framework Programme (FP7/2007-2013)/ERC Grant agreement (291242), Grant-in-Aid for Specially Promoted Research (No. 24000008) of  the Japan Society for the Promotion of Science (JSPS), and Pioneering Project of RIKEN, the Austrian Ministry of Science and Research, Austrian Science Fund (FWF) DK PI (W 1252),  Università di Brescia and Istituto Nazionale di Fisica Nucleare. The authors wish to thank the SMI workshop  and Alexandre Wawrzyniak  for their design and hardware manufacturing.}
}

\titlerunning{ASACUSA Antihydrogen Spectroscopy -- Progress Report}        % if too long for running head

\author{C.~Sauerzopf  \and
A.~Capon \and
M.~Diermaier \and
P.~Dupré \and
Y.~Higashi \and
C.~Kaga \and
B.~Kolbinger \and
M.~Leali\and
S.~Lehner \and
E.~Lodi~Rizzini\and
C.~Malbrunot \and
V.~Mascagna \and
O.~Massiczek \and
D.~J.~Murtagh \and
Y.~Nagata\and
B.~Radics\and
M.~C.~Simon \and
K.~Suzuki \and
M.~Tajima \and
S.~Ulmer\and
S.~Vamosi\and
S.~van~Gorp  \and
J.~Zmeskal \and
H.~Breuker \and
H.~Higaki\and
Y.~Kanai \and
N.~Kuroda \and
Y.~Matsuda \and
L.~Venturelli \and
E.~Widmann \and
Y.~Yamazaki 
}


\institute{C.~Sauerzopf, A.~Capon, M.~Diermaier, B.~Kolbinger, S.~Lehner, C.~Malbrunot, O.~Massiczek, M. C.~Simon, K.~Suzuki, S.~Vamosi, J.~Zmeskal, E.~Widmann\at
              Stefan Meyer Instutute for subatomic physics, Austrian Academy of Sciences, Boltzmanngasse 3, 1090 Vienna, Austria \\
              \email{clemens.sauerzopf@oeaw.ac.at}           %  \\
%             \emph{Present address:} of F.~Author  %  if needed
           \and
           C.~Kaga,  H.~Higaki\at
               Graduate School of Advanced Sciences of Matter, Hiroshima University, Kangamiyama, Higashi-Hiroshima, 739-8530 Hiroshima, Japan
            \and
             P.~Dupré, D. J.~Murtagh, Y.~Nagata, B.~Radics, S.~van Gorp, Y.~Kanai, N.~Kuroda, Y.~Yamazaki \at
               Atomic Physics Laboratory, RIKEN, 2-1 Hirosawa Wako-shi, 351-0198 Saitama, Japan 
            \and
               Y.~Higashi, M.~Tajima, Y.~Matsuda \at
               Institute of Physics, Graduate School of Arts and Sciences, University of Tokyo, 3-8-1 Komaba Meguro-ku, 153-8902 Tokyo, Japan
             \and
               M.~Leali, V.~Mascagna,  E.~Lodi Rizzini, L.~Venturelli \at  
               Dipartimento di Ingegneria dell’ Informazione, Università di Brescia, via Valotti 9, Brescia 25133, Italy and sez. INFN di Pavia
               \and
               C.~Malbrunot, H.~Breuker \at 
               Organisation Européenne pour la Recherche Nucléaire (CERN), 1211 Geneva 23, Switzerland
   \and
               S.~Ulmer \at 
Ulmer Initiative Research Unit, RIKEN
}

\date{Received: date / Accepted: date}

\maketitle

\begin{abstract}
We report the successful commissioning and testing of a dedicated field-ioniser chamber for measuring principal quantum number distributions in antihydrogen as part of the ASACUSA hyperfine spectroscopy apparatus. The new chamber is combined with a beam normalisation detector that consists of plastic scintillators and a retractable passivated implanted planar silicon (PIPS) detector.
\end{abstract}

\section{Introduction}
\label{intro}
The combined symmetry of charge, parity and time (CPT symmetry), is one of the most fundamental requirements in the standard model of particle physics. Up to now no violation of the CPT symmetry has been observed. 

One way of constraining CPT violating parameters in the standard model extension \cite{lehnert} is by comparing spin changing transitions in matter and antimatter \cite{kost2,kostVarg1}. Since the first production of cold antihydrogen \cite{coldAntihydrogen1,coldAntihydrogen2} which is the simplest atomic system built completely of antimatter, experiments were planned to test the CPT invariance in the matter-antimatter regime. 

The ASACUSA collaboration at the CERN Antiproton Decelerator (AD) facility aims to test the CPT invariance by comparing the ground state hyperfine splitting of hydrogen and antihydrogen in a Rabi like experimental setup \cite{rabiHydrogen}.  
In a first step the ASACUSA collaboration already succeeded in producing antihydrogen and proving the feasibility of detecting antihydrogen in a field free region \cite{kuroda2014source}. 

\section{The ASACUSA Antihydrogen Beamline}
\label{sec:2}
The ASACUSA antihydrogen beamline consists of two distinct parts. The first part is the antihydrogen production apparatus which is used to produce a cold and polarised beam of antihydrogen. The second part is the spectroscopy beamline for measuring the ground state hyperfine transitions in antihydrogen.

\subsection{Antihydrogen Production}   
\label{sec:2.1}
Decelerated antiprotons are ejected from the AD into the ASACUSA experimental area. After injection the particles get further decelerated and are captured in a penning type capture trap called MUSASHI \cite{musashi}.  Inside of the so-called double CUSP trap \cite{radic,kuroda2014towards} antihydrogen is formed in a mixing process. The produced antihydrogen is neutral and can therefore escape the trapping fields. By the magnetic cusp field geometry the low-field seeking (LFS) anti atoms (negative magnetic moment) are then focused \cite{nagataLense} towards the spectroscopy beamline. 

\subsection{Spectroscopy Beamline}   
\label{sec:2.2}
\begin{figure}
\centering
\includegraphics[angle=0,width=1\textwidth]{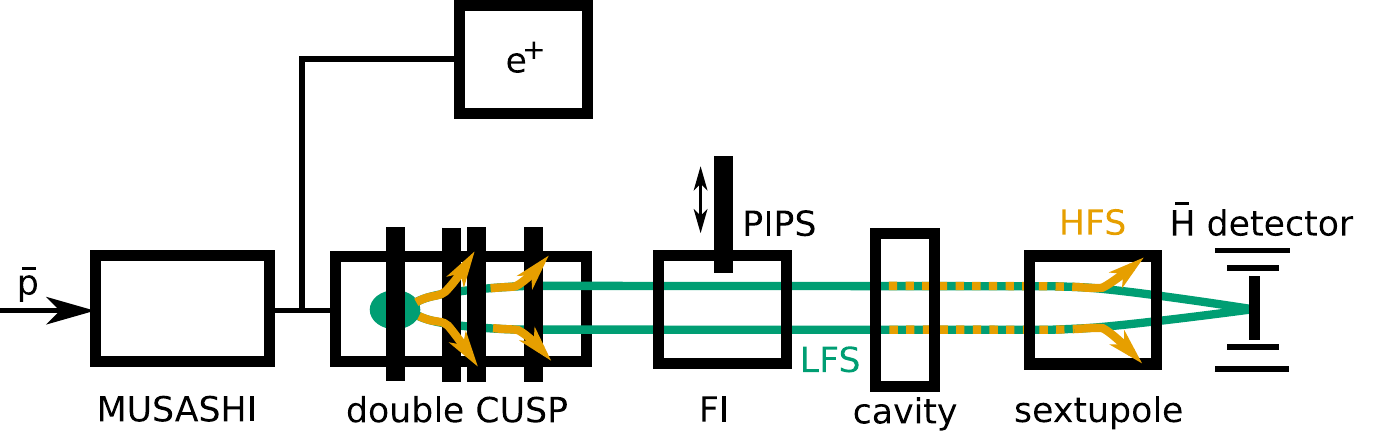}
\caption{Schematic drawing of the ASACUSA hyperfine spectroscopy apparatus. Antiprotons reach the MUSASHI trap from the left side. Antihydrogen is produced in the double CUSP trap and then focused towards the microwave cavity in the case of LFS, or defocused in the case of HFS. If a transition occurs in the cavity, the antiatoms get defocused by the sextupole magnet and annihilate on the beampipe wall. In the off-resonance case, the antihydrogen reaches the detector}
\label{fig:beamline}
\end{figure}

For measuring the transition frequency of ground state antihydrogen the spectroscopy apparatus is the central part of the experiment \cite{Widmann2013,Malbrunot2014hyperfine}. 
The LFS traverse the microwave cavity. If the resonance condition is met a spin flip occurs by the induced hyperfine transition that converts a LFS into a high-field seeking state (HFS). After the microwave cavity the anti atoms reach the superconducting sextupole analyser magnet. If the resonance condition is met, the HFS get defocused by the magnetic field gradient whereas the LFS get focused onto the antihydrogen detector. By tuning the microwave cavity, a resonance scan can be recorded for inferring the ground state hyperfine splitting of antihydrogen. In Fig.~\ref{fig:beamline} a schematic overview is shown.

The effectiveness and performance of the spectroscopy beamline was tested separately with a beam of cold polarised hydrogen \cite{Diermaier2015HyperFine}.

\section{A Combined Field Ioniser and Beam Normalisation Chamber}
\label{sec:3}
The spectroscopy process described above depends on two important conditions. First the antihydrogen atoms have to be in ground state and second the production rate has to be either constant or closely monitored. The production rate, but also the distribution of the principal quantum number \cite{radics2014scaling} are strongly dependent on the mixing conditions in the double CUSP trap.

In order to solve these issues a new vacuum chamber was developed and mounted in-between the microwave cavity and the double CUSP trap. The interior of the chamber contains two planar parallel copper grids with a transparency for antihydrogen of 95\% for each grid. An electric potential difference of 20~kV ($\pm$ 10~kV per grid) can be applied. The grids are mounted with a distance of 10$\pm$0.5~mm. This translates to a minimal principal quantum number n$\ge$12 ($\approx$16~kV/cm) to be ionised when traversing the field ionisation region \cite{kuroda2014source}.

\begin{figure}
\centering
\includegraphics[angle=0,width=.59\textwidth]{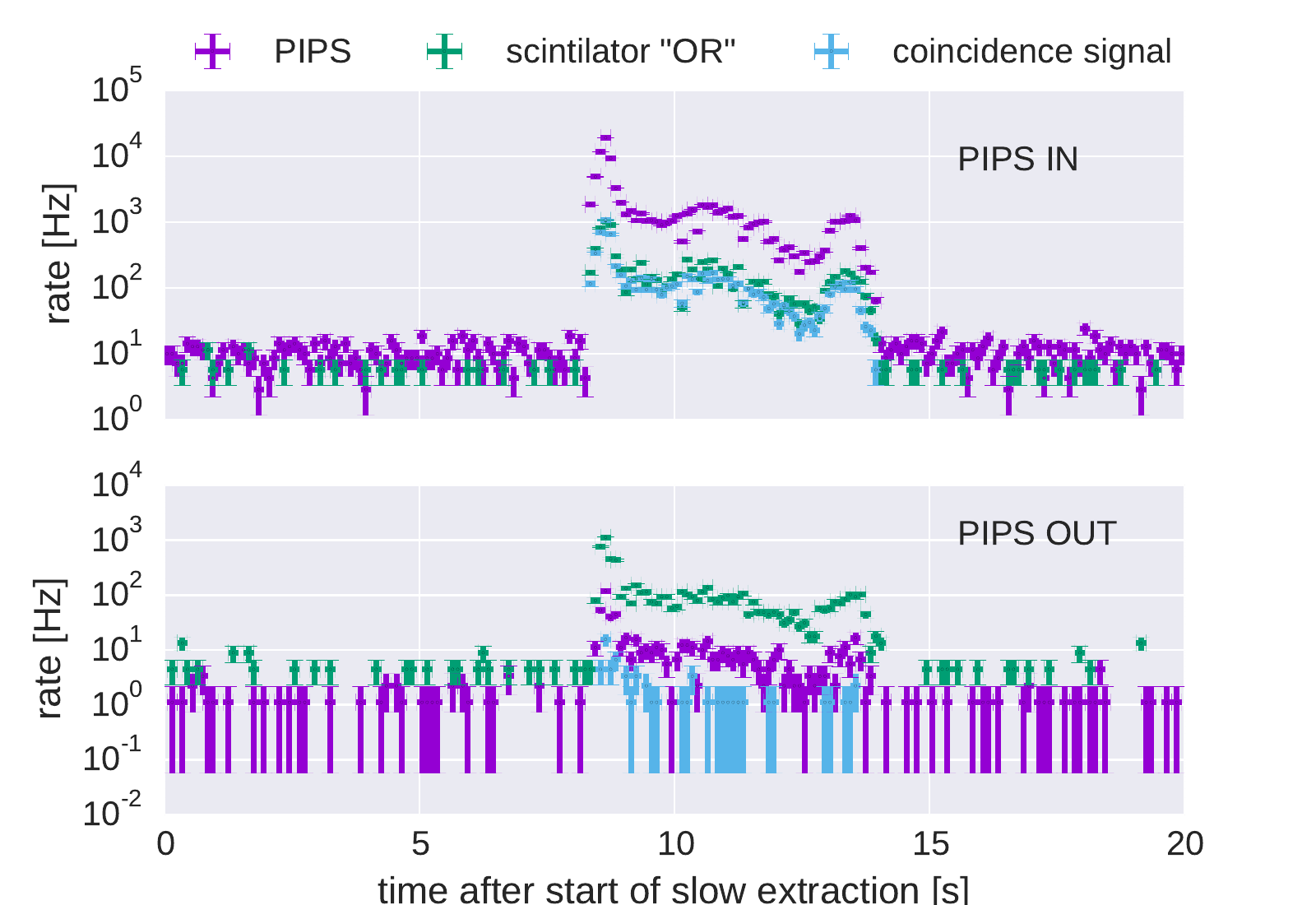}
\raisebox{1.6em}{\includegraphics[angle=90,width=.4\textwidth]{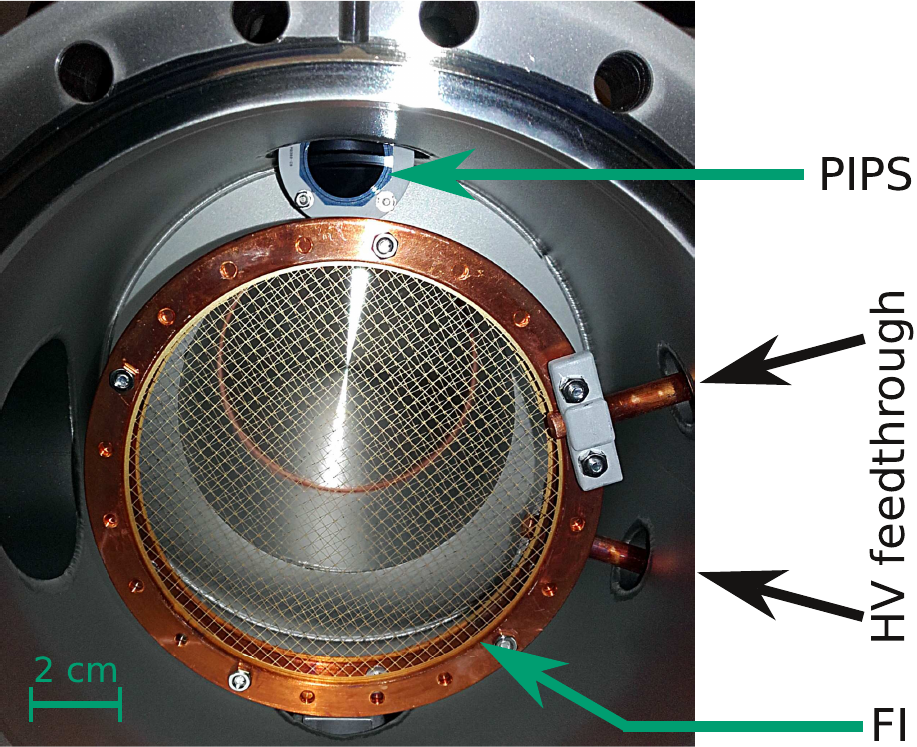}}
\caption{\textbf{Right}: Photograph of the field-ioniser (FI) chamber with high voltage (HV) feedthroughs and the PIPS detector. \textbf{Left}: Time structure of measured count rates during slow extraction of antiprotons from the MUSASHI trap. The rate of the PIPS detector, and the surrounding plastic scintillators alone, and the coincidence between plastic scintillators and PIPS detector is shown. Top: PIPS detector is placed in the beam. Bottom: PIPS detector is retracted from the beam}
\label{fig:pipsData}
\end{figure}

In addition to the field ioniser an active beam blocker with a total diameter of 35~mm, a thickness of 300~$\upmu$m , and with an active surface area of 300~mm\textsuperscript{2} was developed and finally installed in 2014. In general, a beam blocker is required for precision spectroscopy as the superconducting sextupole has a vanishing B field in the centre that would allow HFS to reach the detector and contribute to the background.  The beam blocker is built of a passivated implanted planar silicon (PIPS) detector (Canberra PD300-300CB) that is glued onto a ceramics disc. Together with plastic scintillators surrounding the vacuum chamber the PIPS detector can be user for beam normalisation by taking the coincidence signal between the plastic scintillators and the PIPS detector. In case the field ioniser is used for measurements, the PIPS detector is mounted on a pneumatic actuator that can retract the detector completely from the beam pipe.  A photograph if the chamber components is shown in Fig.~\ref{fig:pipsData}.

\subsection{Antiproton Commissioning}
\label{sec:4}
For commissioning the field ioniser and beam normalisation chamber a slowly extracted beam of antiprotons was created in the MUSASHI trap during the ASACUSA beamtime 2015. The extraction energy was set to 150~eV. A major challenge was imposed by the small opening of the aperture separating the MUSASHI trap from the double CUSP trap. For this reason, the double CUSP magnet was required to stay powered, as a tune from MUSASHI to double CUSP was already optimised for the 150~eV antiprotons. The defocusing effect of the two cusp regions in the magnetic field were counteracted by setting the trap electrodes in these regions to -150~V. 

In Fig. \ref{fig:pipsData} the response of the beam normalisation system is shown. In the top graph the antiprotons could reach the PIPS detector and produce a coincidence signal in the PIPS detector and in the outer plastic scintillators. The bottom graph shows the same measurement but this time the active beam blocker has been retracted from the beam. As a consequence, the antiprotons cannot reach the PIPS detector, and all recorded coincidence events have to be considered as random coincidences. It can be seen that in the latter case almost no coincidence signal is produced. Following this observation, it can be concluded that the PIPS detector and plastic scintillator coincidence system is suitable to serve as a means for beam normalisation. In Table \ref{tab:pipsData} the recorded mean numbers of antiproton annihilations are summarised.

\begin{table}
\caption{Summary of mean number of recorded events in 20 seconds after the start of slow extraction cycle from the MUSASHI trap. Furthermore, the number of extraction cycles is recorded.}
\label{tab:pipsData}
\centering
\begin{tabular}{lcccc}\toprule
 & \multicolumn{3}{c}{mean integrated counts in 20~s} &\\
 & coincidence & PIPS detector & plastic scintillators & \# of runs \\\midrule
PIPS in beam & 756\(\pm\)331 & 9948\(\pm\)4836 & 1021\(\pm\)422 & 7\\
PIPS removed & 6\(\pm\)2 & 80\(\pm\)12 & 706\(\pm\)144 & 9 \\\bottomrule
\end{tabular}
\end{table}

\section{Summary and Outlook}
\label{sec:5}
In 2014 a new vacuum chamber was designed. It houses a  field-ioniser that is capable of ionising antihydrogen atoms with a principal quantum number as low as n=12 and a PIPS detector as active beam blocker that can be used for beam normalisation measurements during data taking.

The new chamber will be used during the upcoming beamtime 2016 for measuring the principle quantum number distribution for antihydrogen produced in the double CUSP trap. The beam normalisation detector is prepared, and was successfully tested for data taking.

\bibliographystyle{spphys}       % APS-like style for physics
\bibliography{biblio.bib}   % name your BibTeX data base

\end{document}